\documentclass[
reprint,aps,amsmath,amssymb,superscriptaddress]{revtex4-1}
\usepackage[utf8]{inputenc}
\usepackage{graphicx}
\usepackage{dcolumn}
\usepackage{bm}
\usepackage[usenames, dvipsnames]{color}
\usepackage{siunitx}
\usepackage[normalem]{ulem} 

\begin{document}


\title{Cavity quantum electro-optics:\\
Microwave-telecom conversion in the quantum ground state}
\author{William Hease}
\email{Authors contributed equally.}
\affiliation{Institute of Science and Technology Austria, Klosterneuburg, Austria}
\author{Alfredo Rueda}
\email{Authors contributed equally.}
\affiliation{Institute of Science and Technology Austria, Klosterneuburg, Austria}
\affiliation{Department of Physics, University of Otago, Dunedin, New Zealand}
\author{Rishabh Sahu}
\affiliation{Institute of Science and Technology Austria, Klosterneuburg, Austria}
\author{Matthias Wulf}
\affiliation{Institute of Science and Technology Austria, Klosterneuburg, Austria}
\author{Georg Arnold}
\affiliation{Institute of Science and Technology Austria, Klosterneuburg, Austria}
\author{Harald G. L. Schwefel}
\affiliation{Department of Physics, University of Otago, Dunedin, New Zealand}
\affiliation{The Dodd-Walls Centre for Photonic and Quantum Technologies, New Zealand}
\author{Johannes M. Fink}
\email{jfink@ist.ac.at}
\affiliation{Institute of Science and Technology Austria, Klosterneuburg, Austria}

\date{\today}
\begin{abstract}
Fiber optic communication is the backbone of our modern information society, offering high bandwidth, low loss, weight, size and cost, as well as an immunity to electromagnetic interference \cite{Hecht1999}. Microwave photonics lends these advantages to electronic sensing and communication systems \cite{Capmany2007}, but - unlike the field of nonlinear optics -
electro-optic devices so far require classical modulation fields whose variance is dominated by electronic or thermal noise rather than quantum fluctuations. Here we present a cavity electro-optic transceiver \cite{Matsko2007,Tsang2010,Tsang2011,Rueda2016,Javerzac-Galy2016,Soltani2017} operating in a millikelvin environment with a mode occupancy as low as $0.025\,\pm0.005$ noise photons. Our system is based on a lithium niobate whispering gallery mode resonator, resonantly coupled to a superconducting microwave cavity via the Pockels effect \cite{Cohen2001,Ilchenko2003,Fan2018,Witmer2020}. For the highest continuous wave pump power of 1.48\,mW we demonstrate bidirectional single-sideband conversion of X band microwave to C band telecom light with a total (internal) efficiency of 0.03\% (0.7\%) and an added output conversion noise of 5.5 photons. The high bandwidth of $10.7\, \text{MHz}$ combined with the observed very slow heating rate of 1.1\,noise photons\,$\text{s}^{-1}$ puts quantum limited pulsed microwave-optics conversion 
within reach.  
The presented device is versatile and compatible with superconducting qubits \cite{Reed2017}, which might open the way for fast and deterministic entanglement distribution between microwave and optical fields \cite{Rueda2019,Zhong2020}, for optically mediated remote entanglement of superconducting qubits \cite{Kurpiers2018}, 
and for new multiplexed cryogenic circuit control and readout strategies \cite{Ruedacombs,Youssefi2020}.
\end{abstract}
\maketitle





The last three decades have witnessed the emergence of a great diversity of controllable quantum systems, 
and superconducting Josephson circuits are one of the most promising candidates for the realization of scalable quantum processors \cite{Arute2019}.
However, quantum states encoded in microwave frequency excitations are very sensitive to thermal noise and electromagnetic interference. Short distance quantum networks could be realized with cryo-cooled transmission lines but longer distances and high density networks require coherent upconversion to shorter wavelength information carriers, ideally compatible with existing near infrared  (1550 nm) fiber optic technology. So far no solution exists to deterministically interconnect remote quantum microwave systems, such as superconducting qubits \cite{Arute2019} and quantum dots or spins in solids~\cite{Burkard2020,Awschalom2018} 
via a room temperature link with sufficient fidelity to build large-scale quantum networks \cite{Kimble2008}. Solving this challenge might not only facilitate a new generation of more power efficient classical communication systems \cite{Capmany2007}, but eventually also enable quantum secure communication, modular quantum computing clusters \cite{Wehner2018} and powerful quantum sensing networks.

An ideal quantum signal converter \cite{Lecocq2016} needs to achieve a total bidirectional conversion efficiency close to unity $\eta\sim1$ for quantum level signals with a minimum amount of added noise $N \ll 1$ over a large instantaneous bandwidth that allows for fast transduction compared to typical qubit coherence times. 
Many different platforms are already being explored for microwave to optical photon conversion \cite{Lambert2020,Lauk2020}. Electro-optomechanical systems have shown very encouraging efficiencies \cite{Higginbotham2018, Arnold2020}, but typically suffer from a limited bandwidth in the kHz range. Electro-optic \cite{Rueda2016,Fan2018,Witmer2020} or piezo-optomechanical \cite{Vainsencher2016, Jiang2020, Han2020} conversion can be faster but the conversion noise properties have not been quantified. Facilitated by efficient photon counting and low duty cycle operation, unidirectional transduction of quantum level signals has also recently been shown \cite{Forsch2020,Mirhosseini2020}, but ground state operation has not been demonstrated in a bidirectional interface so far. 
In this work we present such a device operating continuously with a microwave mode occupancy $N_\text{e} \leq 1$, for a pump laser power of up to $P_p=23.5\,\mu$W resulting in a total bidirectional conversion efficiency of $\eta_\text{tot}=9.1\times10^{-6}$.
%
The maximum achieved total 
efficiency of $\eta_\text{tot}=0.03\,\%$ 
is limited by the highest pump laser power of $P_p=1.48$~mW for the available setup at millikelvin temperatures.\\

\noindent\textbf{Theory}\\
Electro-optic converters make use of the nonlinear properties of non-centrosymmetric crystals to couple optical and microwave degrees of freedom. Our resonant transducer  has two high quality factor optical modes whose frequency difference matches the resonance frequency of a microwave mode. The system's interaction Hamiltonian is given as~\cite{Tsang2010}
\begin{equation} \label{hamsqu}
\hat{H}_\text{int}=\hbar g (\hat{a}_e\hat{a}_p\hat{a}_o^\dagger+\hat{a}_e^\dagger\hat{a}_p^\dagger\hat{a}_o),
\end{equation}
where $\hat{a}_e$, $\hat{a}_p$, and $\hat{a}_o$ stand for the annihilation operators for the microwave, optical pump, and optical signal mode, respectively. This Hamiltonian describes two reciprocal three-wave mixing processes that involve creation and annihilation of photons while respecting energy conservation. The nonlinear vacuum coupling rate $g$ for this interaction depends on the material's effective electro-optic coefficient $r$ and the spatial overlap of the three modes \cite{Rueda2016}
\begin{equation} \label{g_formula}
g= r\, \sqrt{\frac{\varepsilon_p \varepsilon_o}{\varepsilon_e}}  \sqrt{\frac{\hbar\omega_e\omega_p\omega_o}{8\varepsilon_0V_eV_pV_o}} \int dV \psi_e \psi_p \psi_o^\star,
\end{equation} 
with the mode frequency $\omega_k$, the relative permittivity $\varepsilon_k$ and permeability $\mu_k=1$, the effective mode volume $V_k$, and the normalized spatial field distribution $\psi_k$ defined such that the single-photon electric field for mode $k \in \{e,p,o\}$ can be written as $E_k(\vec{r}) = \sqrt{\hbar\omega_k / (2\varepsilon_0\varepsilon_k V_k)} \psi_k(\vec{r})$.  All three modes are whispering gallery modes (WGM) \cite{strekalov2016}
whose spatial field distribution can be separated in the cross-sectional  
and azimuthal part 
$\psi_k(r, z, \phi) = \Psi_k(r, z) e^{-i m_k \phi}$. The integral in Eq.~(\ref{g_formula}) is non-zero only if the azimuthal numbers of the participating modes fulfill $m_o=m_p+m_e$, which is also known as phase matching or angular momentum conservation.

In our conversion scheme we drive the mode $\hat{a}_p$ with a bright coherent tone $\hat{a}_p\rightarrow \alpha_p$, which simplifies   
Eq.~(\ref{hamsqu}) to
\begin{equation} \label{hamsqulin}
\hat{H}_\text{int}= \hbar \alpha_p g  (\hat{a}_e \hat{a}_o^\dagger+\hat{a}_e^\dagger\hat{a}_o).
\end{equation}
This is known as the beam splitter Hamiltonian and it corresponds to a linear coupling between the optical mode $\hat{a}_o$ and microwave  mode $\hat{a}_e$. From the enhanced coupling rate $\alpha_p g$, we define the multi-photon cooperativity as $C = 4 |\alpha_p|^2g^2 / (\kappa_o\kappa_e)$, where $\kappa_o$ and $\kappa_e$ are the total loss rates of the optical and microwave mode, respectively. The multi-photon cooperativity is the figure of merit in most of the resonant electro-optic devices, both for frequency conversion and  entanglement generation~\cite{Tsang2011,Rueda2016,Fan2018,Rueda2019}.\\

\begin{figure}[t]
	\centering
		\includegraphics[width=1.0\columnwidth] {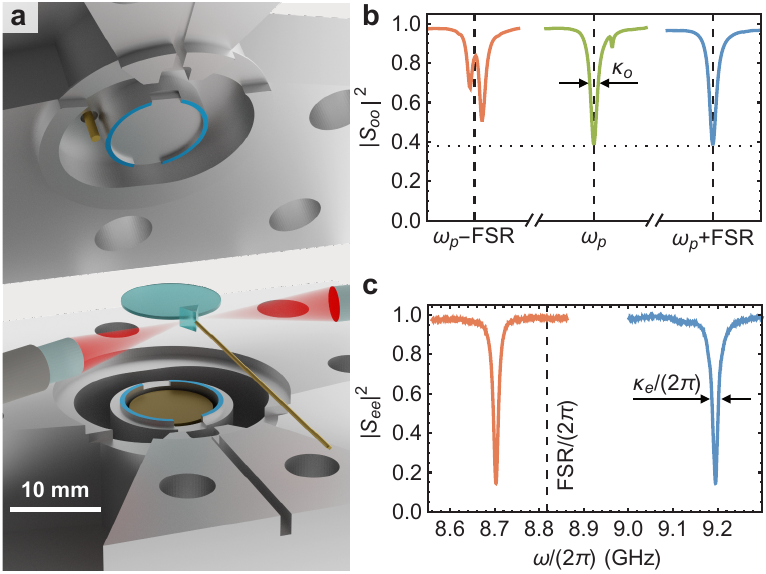}
	\caption{
\textbf{Resonant electro-optic device. a}, Exploded-view rendering of the electro-optic converter. The WGM resonator (light blue disc) is clamped between two aluminum rings (blue shaded areas) of the top and bottom part of the aluminum microwave cavity. Two GRIN-lenses are used to focus the optical input and output beams (red) on a diamond prism surface in close proximity of the optical resonator. The microwave tone is coupled in and out of the cavity using a coaxial pin coupler at the top of the cavity (gold). The prism, both lenses and the microwave tuning cylinder (gold shaded inside the lower ring) positions can be adjusted with 8 linear piezo positioners.
\textbf{b}, Optical reflection spectrum of the WGM resonator at base temperature ($\sim 7$ mK). The optical pump mode at $\omega_p/(2\pi)\approx193.5$\,THz (green) and signal mode (blue) are critically coupled and separated by one free spectral range (FSR, dashed lines). On resonance 38\% of the optical power is reflected without entering the WGM resonator due to imperfect optical mode overlap $\Lambda$ (horizontal dotted line). The lower sideband mode (red) was chosen to couple to a mode family of different polarization, which splits it and facilitates the single-sideband selectivity of the converter.
\textbf{c}, Reflection spectrum of the microwave cavity at base temperature ($\sim 7$ mK) for the tuning cylinder in its up position (blue line) and in its down position (red line).  With a tuning range of $\sim\,0.5$~GHz we can readily match the cavity frequency with that of the optical free spectral range $\text{FSR}/(2\pi) = 8.818$~GHz (dashed line).}
	\label{figure_1}
\end{figure}

\noindent\textbf{Device}\\
The electro-optic transducer consists of a z-cut LiNbO$_3$ WGM resonator, with a major radius $R=2.5$\,mm, sidewall surface radius $\rho\approx0.7$\,mm and a thickness $d=0.15$ mm, coupled to a superconducting aluminum cavity as shown in Fig.~\ref{figure_1}a. 
The top and bottom rings of the cavity 
are designed to confine the microwave mode at the rim of the WGM resonator and maximize the spatial mode overlap with the two optical modes. Here we use type-0 frequency conversion, where all the participating waves are polarized parallel to the material's optic and symmetry axis, addressing the highest electro-optic tensor component of LiNbO$_3$.
We work with two optical modes of the WGM resonator that are spectrally separated by the resonator's optical free spectral range (FSR) as shown in Fig.~\ref{figure_1}b. The pump mode has an azimuthal number $m_p\approx20\times10^3$ and the signal mode $m_o=m_p+1$. The $m_p-1$ mode's participation in the resonant interaction is suppressed due to its  avoided crossing with another mode family \cite{Rueda2016}, 
leaving only 2 optical modes in the process.  
We use an antireflection coated diamond prism to feed the optical pump into the optical resonator via evanescent coupling. The prism is attached to a linear piezo positioning stage 
that allows to accurately tune the extrinsic optical coupling rate $\kappa_{\text{ex},o}$ in-situ. The continuous wave optical pump is a $\sim10$~kHz linewidth coherent laser tone 
that is locked to the resonance of the optical pump WGM at $\omega_p/(2\pi)\approx 193.5$ THz
for conversion measurements. The cryostat optical input line consists of a single mode fiber with a GRIN-lens at its end to focus the optical beam at the prism-WGM resonator coupling point. The reflected optical pump is collected with a second GRIN-lens and coupled to the output line fiber for further measurements at room temperature. 

At base temperature ($\sim 7$~mK) we measure an optical mode separation $\text{FSR}/(2\pi)=8.818$~GHz and an intrinsic optical loss rate $\kappa_{\text{in},o}=9.46$~MHz, which corresponds to a quality factor $Q_{\text{in},o}=2.0\times10^7$ - a reduction 
by a factor 10 (5) from the measured room temperature value outside (inside) the microwave cavity. The chosen optical pump and signal modes have a contrast of 
62\% at critical coupling ($Q_{\text{ex},o}=Q_{\text{in},o}$) as shown in Fig.~\ref{figure_1}b, due to an imperfect spatial field mode overlap $\Lambda^2=0.38$ between the optical WGM and the optical input beam
(see Supplementary Material). In this work we keep the optical system critically coupled to maximize the optical photon number for a given input power. The optical signal at $\omega_o=\omega_p+\text{FSR}$ for optical to microwave conversion is created using a suppressed-carrier single-sideband modulator and sent through the same optical path as the pump tone, see Supplementary Information.

The chosen microwave cavity mode undergoes one oscillation around a full azimuthal roundtrip $m_e=1$, and its frequency is matched to the optical FSR in order to fulfill the conditions of phase matching and energy conservation. We use an aluminum cylinder centered below the WGM resonator and attached to a vertical piezo positioner that shifts the microwave resonance frequency $\omega_e/(2\pi)$ from 8.70 to 9.19 GHz at base temperature as shown in Fig.~\ref{figure_1}c. 
Microwave tones are sent to the device through a heavily attenuated transmission line and subsequently coupled to the cavity via a coaxial pin coupler mounted in the top part of the cavity as shown in Fig.~\ref{figure_1}a. The reflected microwave tone and the down-converted optical signal pass two circulators before amplification and measurement with a vector network analyzer (VNA) or an electronic spectrum analyzer (ESA), see Supplementary Information. From the VNA reflection measurements, we extract the resonance frequency $\omega_e$, the intrinsic loss rate $\kappa_{\text{in},e}=6.7$ MHz 
and the extrinsic coupling rate $\kappa_{\text{ex},e}=3.7$ MHz of the microwave resonance mode.\\

\noindent\textbf{Bidirectional Conversion}\\
In our system the microwave-to-optics  
and optics-to-microwave 
photon conversion efficiencies are equal \cite{Tsang2011}. The total input-output electro-optic photon conversion efficiency, defined on resonance, is given as 
\begin{equation}
\eta_\text{tot}= \eta_e\eta_o\Lambda^2  \frac{4C}{(1+C)^2}\label{conversion_eff},
\end{equation}
with coupling efficiencies $\eta_k = \kappa_{\text{ex},k} / \kappa_k$ and $\kappa_k= \kappa_{\text{in},k}+ \kappa_{\text{ex},k}$.  
We determine the bidirectional conversion efficiency of the device $\eta_\text{tot}=\sqrt{\eta_{eo}\eta_{oe}}$, independent of the specifics of the measurement setup \cite{Andrews2014}, such as the optical and microwave input attenuations $\beta_1, \beta_3$ and output amplifications $\beta_2 ,\beta_4$ (see Supplementary Material). Performing 4 independent measurements of the coherent scattering parameters 
$|S_{ij}|^2\propto|\hat{a}_{\text{out},i}/\hat{a}_{\text{in},j}|^2$ 
with $i,j=\{e,o\}$ for every optical pump power setting
\begin{equation} \label{conversion_autonorm}
\eta_\text{tot} =\sqrt{\frac{|S_{eo}(\omega_0)|^2\cdot|S_{oe}(\omega_0)|^2}{|S_{ee}(\omega_\Delta)|^2\cdot|S_{oo}(\omega_\Delta)|^2}}= \sqrt{\frac{\beta_1 \eta_{eo} \beta_4 \cdot \beta_3 \eta_{oe} \beta_2}{\beta_1 \beta_2 \cdot \beta_3 \beta_4}},
\end{equation}
we obtain the in-situ calibrated device efficiency $\eta_\text{tot}$ from the optical fiber to the microwave coaxial line. Here the optics-to-microwave $|S_{eo}|^2$ and microwave-to-optics $|S_{oe}|^2$ power ratios are measured on resonance $\omega_0=\omega_e, \omega_o$ and the reflected optical $|\text{S}_{oo}|^2$ and microwave $|\text{S}_{ee}|^2$ tones are measured at a detuning such that $|\omega_\Delta-\omega_0|\gg\kappa_e,\kappa_o$ respectively. For higher accuracy we take into account frequency dependent baseline variations using the full measured reflection scattering parameters.

\begin{figure}[t!]
	\centering
		\includegraphics[width=0.99\columnwidth]{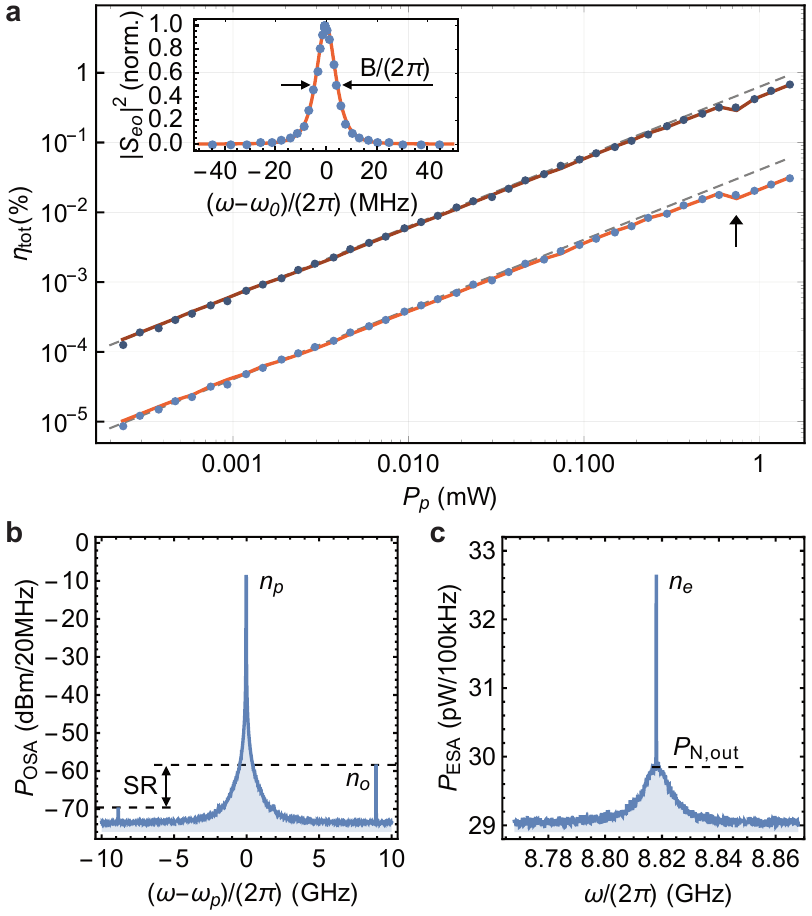}
\caption{
\textbf{Bidirectional microwave-optics conversion. a},~Measured photon conversion efficiency $\eta_\text{tot}$ (light blue points) and inferred internal device efficiency $\eta_\text{int}=\eta_\text{tot}/(\eta_e\eta_o\Lambda^2)$ (dark blue points) together with theory (red lines), i.e. Eq.~(\ref{conversion_eff}) taking into account measured cavity linewidth changes.
The dashed lines are linear fits for the 10 lowest power data points respectively. The arrow marks the input power 
where the aluminum cavity goes from super to normal conducting. The inset shows the measured and normalized coherent optics-to-microwave conversion power ratio for $P_p=18.7\,\mu\text{W}$ and $P_o=267\,\text{nW}$, as a function of the detuning between the optical signal frequency and $\omega_o$ (blue points) together with theory Eq.~(\ref{Bandwidth}) (red line), indicating the conversion bandwidth $B/(2\pi)=9.0$~MHz.
\textbf{b},~Measured optical power spectrum for microwave-to-optical conversion at $P_p=1.48\,\text{mW}$. 
The weak coherent microwave tone $P_e=1.0\,\text{nW}$ 
generates two optical sidebands (blue and red) with a suppression ratio of $SR=10.7$~dB. 
The center and sideband peaks are proportional to the intra-cavity pump $n_p$ and converted optical signal $n_o$ photon numbers respectively. The noise floor is set by the resolution bandwidth.
\textbf{c},~Measured power spectrum for optical-to-microwave conversion at $P_p=2.35\,\mu\text{W}$. 
The weak optical input signal $P_o=161\,\text{nW}$ generates a single coherent microwave tone at $\omega_o-\omega_p$. In this particular example $n_e=1.2$ intra-cavity microwave photons are generated with an incoherent noise floor $P_{N,\text{out}}$ corresponding to an added output conversion noise of $N_\text{out}=0.4\,\text{photons}\,\text{s}^{-1}\,\text{Hz}^{-1}$ in the center of the microwave cavity bandwidth.
}
\label{figure_2}
\end{figure}

In Fig.~\ref{figure_2}a we show the measured values of the total $\eta_\text{tot}$ (light blue) and the calculated internal conversion efficiency $\eta_\text{int}=\eta_\text{tot}/(\eta_e\eta_o\Lambda^2)$ (dark blue) together with Eq.~(\ref{conversion_eff}) taking into account measured cavity linewidth changes (red lines) as a function of the incident optical pump $P_p$.  As the pump power increases, the conversion efficiency departs only slightly from the expected linear behavior for the low cooperativity limit $(C\ll1)$ (dashed lines). 
For $P_p \approx$ 700 $\mu$W (arrow), $\eta_\text{tot}$ drops because the aluminum cavity undergoes a phase transition from the superconducting to the normal conducting state, which is accompanied by a sudden increase of $\kappa_{\text{in},e}$, see Supplementary Information. The highest conversion efficiency $\eta_\text{tot} = 3.16\times10^{-4}$ is reached for the maximum available pump power $P_p =1.48$ mW, where the refrigerator base plate reaches a steady state temperature of $T_f=320$~mK with theoretical microwave mode occupation of $N_f=0.36$. 

From the measured values of the bidirectional conversion efficiency $\eta_\text{tot}$ and coupling rates $\kappa_{\text{in},e}$ 
at each optical pump power $P_p$, which is related to the drive strength and pump photon number $n_p=|\alpha|^2=4 P_p \Lambda^2 \kappa_{\text{ex},o}/(\hbar\omega_p\kappa_o^2)$, we extract the values of the multi-photon cooperativity in the system, ranging from $C=1.23\times10^{-7}$ for the lowest, to $C=1.67\times10^{-3}$ for the highest $P_p$. From this we deduce the maximum internal photon conversion efficiency $\eta_\text{int}=4C/(1+C)^2$ of 0.67\%. 
We find very good agreement between the measured conversion efficiency and Eq.~(\ref{conversion_eff}) (solid red lines) for $g/(2\pi)=40$~Hz, close to the directly measured (simulated) value of 36.1 Hz (36.2 Hz) at room temperature.


The normalized optics-to-microwave conversion as a function of the optical signal frequency
is shown in the inset of Fig.~\ref{figure_2}a. The solid red line corresponds to the theoretical expectation for the conversion spectrum \cite{Tsang2011}
\begin{multline}
\frac{|\text{S}_{ij}(\omega-\omega_{0})|^2}{|\text{S}_{ij} (\omega_{0})|^2}=\\
\left(\left(1-\frac{4(\omega-\omega_{0})^2}{\kappa_o\kappa_e}\right)^2   +\frac{4(\omega-\omega_{0})^2(\kappa_o+\kappa_e)^2}{\kappa^2_o\kappa^2_e}    \right)^{-1},
\label{Bandwidth}
\end{multline}
where $\kappa_e$ and $\kappa_o/(2\pi)=18.92$~MHz were independently extracted from direct reflection measurements. The bandwidth $B/(2\pi) = 9.0$\,MHz at $P_p=18.7\,\mu\text{W}$ with $\kappa_e/(2\pi)=11.32$ MHz is in excellent agreement with the theoretical model for both conversion directions (see Supplementary Material). $B/(2\pi)$ increases from 8.51\,MHz (calculated, $\kappa_e=10.45$ MHz) for the lowest to 10.68\,MHz (measured, $\kappa_e=14.85$ MHz) for the highest optical pump power.


Selective up-conversion is an important feature of electro-optic transducers, because of the intrinsic noiseless nature of the up-conversion process. Figure \ref{figure_2}b displays the measured microwave-to-optics conversion power spectrum corresponding to the highest pump power. Single sideband conversion with a suppression ratio of 10.7~dB in favor of up-conversion can be observed. This is expected from the asymmetric FSR in our WGM resonator due to the splitting of the lower sideband mode as shown in Fig.~\ref{figure_1}b. The generated microwave output power spectrum from the optics-to-microwave conversion 
is shown in Fig.~\ref{figure_2}c, where the peak at the center represents the coherently converted signal power at the microwave cavity output and the broadband incoherent baseline is due to the thermal noise added to the microwave output as a result of optical absorption.\\

\noindent\textbf{Added noise}\\
The optical pump causes dielectric heating due to absorption in the lithium niobate. In addition, stray light and evanescent fields can lead to direct breaking of Cooper pairs in the superconducting cavity. Both effects cause an increased surface resistance and in turn a larger microwave cavity linewidth $\kappa_{\text{in},e}$ (see Supplementary Information). The optical heating causes an increase of the microwave cavity bath $N_b$ and the microwave waveguide bath $N_\text{wg}$, both are related to the incoherent microwave mode occupancy  \cite{Xu2020}
\begin{equation}
N_e=\eta_e N_\text{wg}+(1-\eta_e)N_b.
\end{equation}
The two bath populations are directly accessible via the detected output noise spectrum given by 
\begin{equation}
N_\text{det}(\omega)=  
\frac{4\kappa_{\text{in},e}\kappa_{\text{ex},e}}{\kappa^2_e+4\omega^2}\left(N_{b}-N_\text{wg}\right)  + N_\text{wg} + N_\text{sys}
\label{noiseeq}
\end{equation}
in the low cooperativity limit. The conversion noise at the output port of the device $N_\text{out}=N_\text{det}-N_\text{sys}$ in units of $\text{photons}\,\text{s}^{-1}\,\text{Hz}^{-1}$ is related to the measured power spectrum $P_\text{ESA}$ via $N_\text{det}(\omega)= P_\text{ESA}(\omega)/(\hbar \omega_e \beta_4)$. 
Here $N_\text{sys}=
12.74\pm0.36$ and $\beta_4=(67.05\pm0.16)$\,dB are the calibrated noise photon number and gain of the measurement setup as referenced to the converter output port. 

\begin{figure*}[t!]
	\centering
		\includegraphics[width=\textwidth]{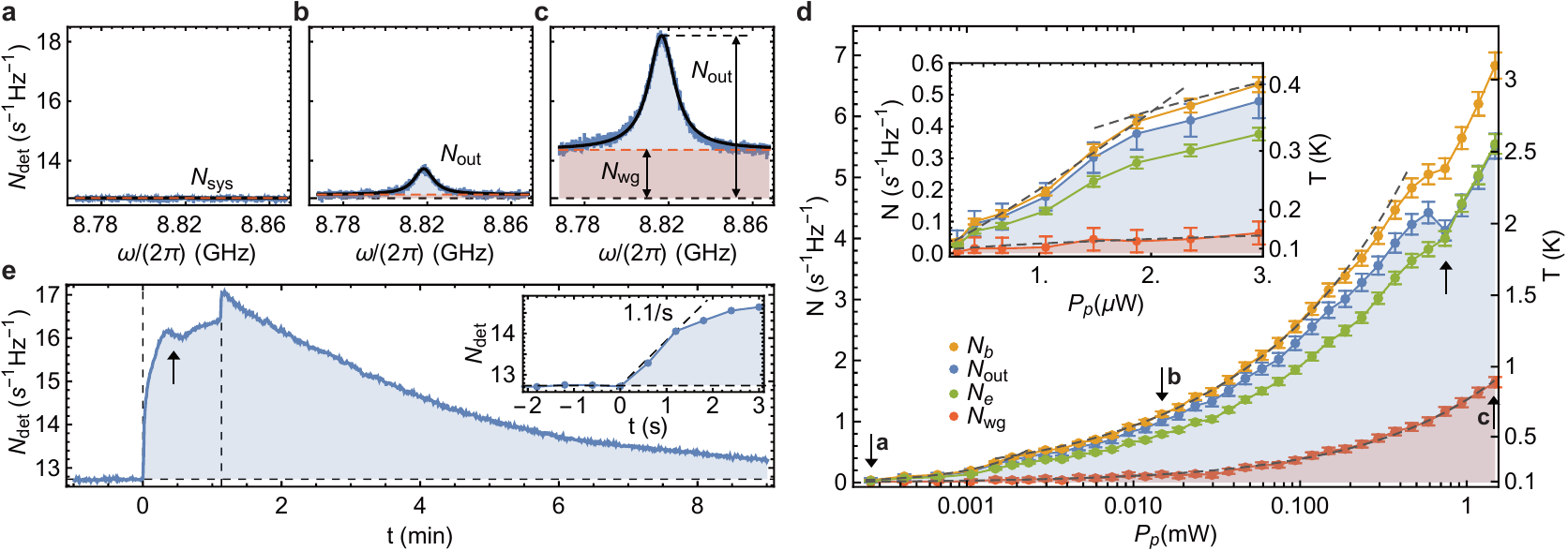}	
\caption{ 
\textbf{Conversion noise and mode population. a, b, c,}~Measured microwave output noise spectrum (blue line) in units of $\text{photons}\,\text{s}^{-1}\,\text{Hz}^{-1}$ for (a) $P_p=0.23\,\mu$W, (b) 14.82\,$\mu$W and (c) 1.48\,mW together with a fit to Eq.\,(\ref{noiseeq}) (black line). In all three panels the dashed black line indicates the measurement system noise floor $N_\text{sys}$ and the dashed red line indicates the broad-band noise offset $N_\text{wg}$. 
\textbf{d,}~External waveguide bath population $N_\text{wg}$ (red), total output noise photons $N_\text{out}$ (blue), microwave mode bath $N_b$ (yellow) and mode occupancy $N_e$ (green) as a function of $P_p$. The error bars of $N_\text{wg}$ and $N_\text{out}$ at low $P_p$ are dominated by systematic errors due to slow absolute variations of the baseline of $\pm0.03$ quanta. The error bars of $N_b$ represent the 95\% confidence interval of the fit to Eq.\,(\ref{noiseeq}) which also dominates the uncertainty of $N_e$ since $N_\text{wg}<N_\text{b}$. The error bars at high $P_p$ are dominated by the accuracy of the $N_\text{sys}$ calibration. The shown error bars are the extrema of these absolute and relative uncertainties.
The inset shows the region where the microwave bath occupancy $N_b < 1$ on a linear scale. The dashed gray lines indicate fitted power laws, specifically $N_\text{wg}\propto P_p^{0.55}$ over the full range of powers, $N_\text{b}\propto P_p^{1.14}$ up to $P_p\sim2\,\mu$W (see inset) and $N_\text{b}\propto P_p^{0.45}$ at higher powers.
\textbf{e,}~Microwave output noise $N_\text{out}$ measured on resonance over a 500\,kHz resolution bandwidth in units of $\text{photons}\,\text{s}^{-1}\,\text{Hz}^{-1}$ (blue line) on top of the measurement system noise floor $N_\text{sys}$ (dashed horizontal line) as a function of time. The system is excited with a rectangular optical pulse of $P_p=1.48\,$mW and a length of 68 seconds (dashed vertical lines). The arrow indicates the time of approximately 30 seconds when superconductivity breaks, the cavity quality factor degrades and the frequency detunes from the measurement frequency. This process is reversed at the end of the pulse when the cavity tunes back and the detected noise increases temporarily. The inset shows the fastest time scale, i.e. the initial heating rate of $dN_\text{out}/dt\approx1.1\,\text{noise photons\, s}^{-1}$.
}\label{fig_3}
\end{figure*} 

In Fig.~\ref{fig_3}a-c we show the measured noise spectrum obtained by normalizing with a no-pump baseline reference measurement when the sample is cold $N_\text{det}=N_\text{sys} P_\text{ESA}/P_{\text{ESA}(P_p=0)}$ for three different pump powers with the same y-axis scale and no signal tone applied. For the lowest pump power $P_p = 0.23\, \mu$W only the $N_\text{sys}$ offset is discernible (dashed black lines in panels a-c). For the intermediate power $P_p=14.82\, \mu$W the total output microwave noise $N_\text{out}=1.01\pm0.07$ appears as a Lorentzian curve (blue line) with a broad band noise background $N_\text{wg}=0.13\pm0.04$ (red dashed line). For the largest applied power $P_p=1.48$\,mW we observe a maximum of $N_\text{out}=5.51\pm0.20$ and $N_\text{wg}=1.64\pm0.08$, significantly hotter than the dilution refrigerator base plate at $N_f=0.36$. This is expected for a steady-state localized noise source, such as the optically pumped dielectric resonator, which has a finite temperature dependent thermalization rate to equilibrate with the environment.




The added conversion noise referenced to the device output $N_\text{out}$ (blue), the broad band waveguide noise $N_\text{wg}$ (red), the microwave bath $N_b$ (yellow) and mode occupancy $N_e$ (green) for different optical pump powers $P_p$ are shown in Fig.~\ref{fig_3}d. Sub-photon microwave output noise as low as $N_\text{out}=0.03^{+0.04}_{-0.03}$ and microwave mode occupancies as low as $N_e=0.025\pm0.005$ are achieved for a continuous wave pump power of $P_p = 0.26\, \mu$W where the total conversion efficiency is $\eta_\text{tot}=8.8\times10^{-8}$.

As the pump power is increased, we observe a smooth growth of the waveguide noise starting from an equivalent temperature of $T_\text{wg}=78^{+50}_{-17}$\,mK and roughly proportional to $\sqrt{P_p}$ over 4 orders of magnitude. This is expected if the effective thermal conductivity to the cold refrigerator bath of approximately constant temperature is increasing linearly such that the heat flow $q$ matches the dissipated part of the pump power $q \propto P_p \propto T \cdot \Delta T$, as predicted \cite{Woodcraft2005} for normal conducting metals such as the copper coaxial port attached to the superconducting cavity.  

In contrast, for the microwave bath we observe 3 distinct regions of heating. Up to about $2\,\mu$W the scaling is approximately linear, which is expected for local heating with a fixed thermalization to the cold bath. The thermal conductivity of superconducting aluminum far below the critical temperature is exponentially suppressed \cite{Woodcraft2005} so this thermalization could be due to radiation or direct excitation of quasiparticles. In this important range of noise photon numbers, a high conductivity copper cavity might therefore show a significantly slower trend. Above $2\,\mu$W the scaling is approximately $\sqrt{P_p}$, which indicates that part of the cavity, such as the small rings holding the disk, are normal conducting. This is confirmed by an increase in the internal losses (see Supplementary Information). At $P_p\approx0.7$\,mW we see a sharp drop in the output noise due to a sudden increase of $\kappa_{\text{in},e}$ from 8.6 to 11.2\,MHz. The temporarily slower increase of $N_b$ suggests that this is also accompanied by an higher thermalization rate to the cold refrigerator bath, indicating that the entire aluminum cavity undergoes a phase transition at this input power. This interpretation of the data is backed up by stable cavity properties beyond this power (see Supplementary Material). 
The lowest measured bath occupancies are consistent with qubit measurements for a similar amount of shielding without optics \cite{Fink2010} and could be further improved with sensitive radiometry measurements \cite{Wang2019,Scigliuzzo2020}. 

In Fig.~\ref{fig_3}e we show the time dependence of the measured output noise when the system is excited with a resonant optical square pulse. The measured rise time to the maximum power of $P_p=1.48$\,mW is 1\,ms. Facilitated by its macroscopic device design with a large heat capacity and contact surface area to the cold refrigerator bath, we observe that the fastest timescale at the onset of the square pulse is as low as 1.1 photons\,s$^{-1}$. This is roughly $10^7$ times slower compared to state of the art microscopic microwave devices pulsed with $\sim10^3$ times lower power 
\cite{Mirhosseini2020}. Assuming - as a worst case scenario - a linear increase of the heating rate with the applied power, we can project $N_\text{out}<10^{-4}$ for a single 100\,ns long pulse of power 1\,W.
For this power $C>1$ with unity internal conversion efficiency and interesting new physics to be unlocked.\\






\noindent\textbf{Conclusion}\\
The presented bidirectional microwave-optical interface operates in the quantum ground state $N_e\ll1$, as verified by measuring the minimal noise $N_\text{out}\ll1$ added to a converted microwave output signal. Compared to recent probabilistic unidirectional transduction of quantum level signals we showed somewhat lower \cite{Mirhosseini2020} and orders of magnitude higher \cite{Forsch2020} efficiency. The very high instantaneous bandwidth of $10.7$\,MHz compared to typical $100$\,Hz repetition rates in previous experiments provides a very promising outlook to be able to also verify the quantum statistics using sensitive heterodyne \cite{Rueda2019}
or photon detection measurements \cite{Zhong2020} in the near future. 
Furthermore, bandwidth-matched high power pulsed operation schemes should also enable deterministic protocols due to the observed slow heating timescales, i.e. the conversion of quantum level signals with an equivalent input noise $N_\text{in}= N_\text{out}/\eta_\text{tot}\ll 1$. Such a fast and high fidelity quantum microwave photonic interface together with the non-Gaussian resources of superconducting qubits \cite{Kurpiers2018} might then provide the practical foundation to extend the range of current fiber optic quantum networks \cite{Briegel1998} in analogy to optical-electrical-optical repeaters in the early days of classical fiber optic communication \cite{Hecht1999}.

\textbf{Acknowledgements}
The authors acknowledge the support from T. Menner, A. Arslani, and T. Asenov from the Miba machine shop for machining the microwave cavity, and S. Barzanjeh, F. Sedlmeir and C. Marquardt for fruitful discussions. This work was supported by IST Austria, the European Research Council under grant agreement number 758053 (ERC StG QUNNECT), and the Austrian Science Fund (FWF) through BeyondC (F71). WH is the recipient of an ISTplus postdoctoral fellowship with funding from the European Union's Horizon 2020 research and innovation program under the Marie Sk\l{}odowska-Curie grant agreement No. 754411. GA is the recipient of a DOC fellowship of the Austrian Academy of Sciences at IST Austria. JMF acknowledges support from the European Union's Horizon 2020 research and innovation programs under grant agreement No 732894 (FET Proactive HOT), 862644 (FET Open QUARTET), and a NOMIS foundation research grant.\\
\textbf{Authors contributions}
WH, AR, RS, MW and GA worked on the setup and performed the measurements. Data analysis was done by WH, AR and JMF. Analytical modeling and analysis was done by AR and FEM simulations by RS. AR, HGLS and JMF conceived the project. All authors contributed to the manuscript. JMF supervised the project.\\
\textbf{Competing interests} 
The authors declare no competing interests.\\
\textbf{Data availability} 
The data and code used to produce the results of this manuscript will be made available via an online repository before publication.

\bibliographystyle{naturemag_noURL}
\bibliography{FinkGroupBib_v7}


\newpage
\onecolumngrid
\newpage
\renewcommand{\thefigure}{S\arabic{figure}}
\appendix
\tableofcontents
\setcounter{figure}{0} 


\section{Measurement setup}
\begin{figure}[h]
	\centering
		\includegraphics[width=0.8\textwidth]{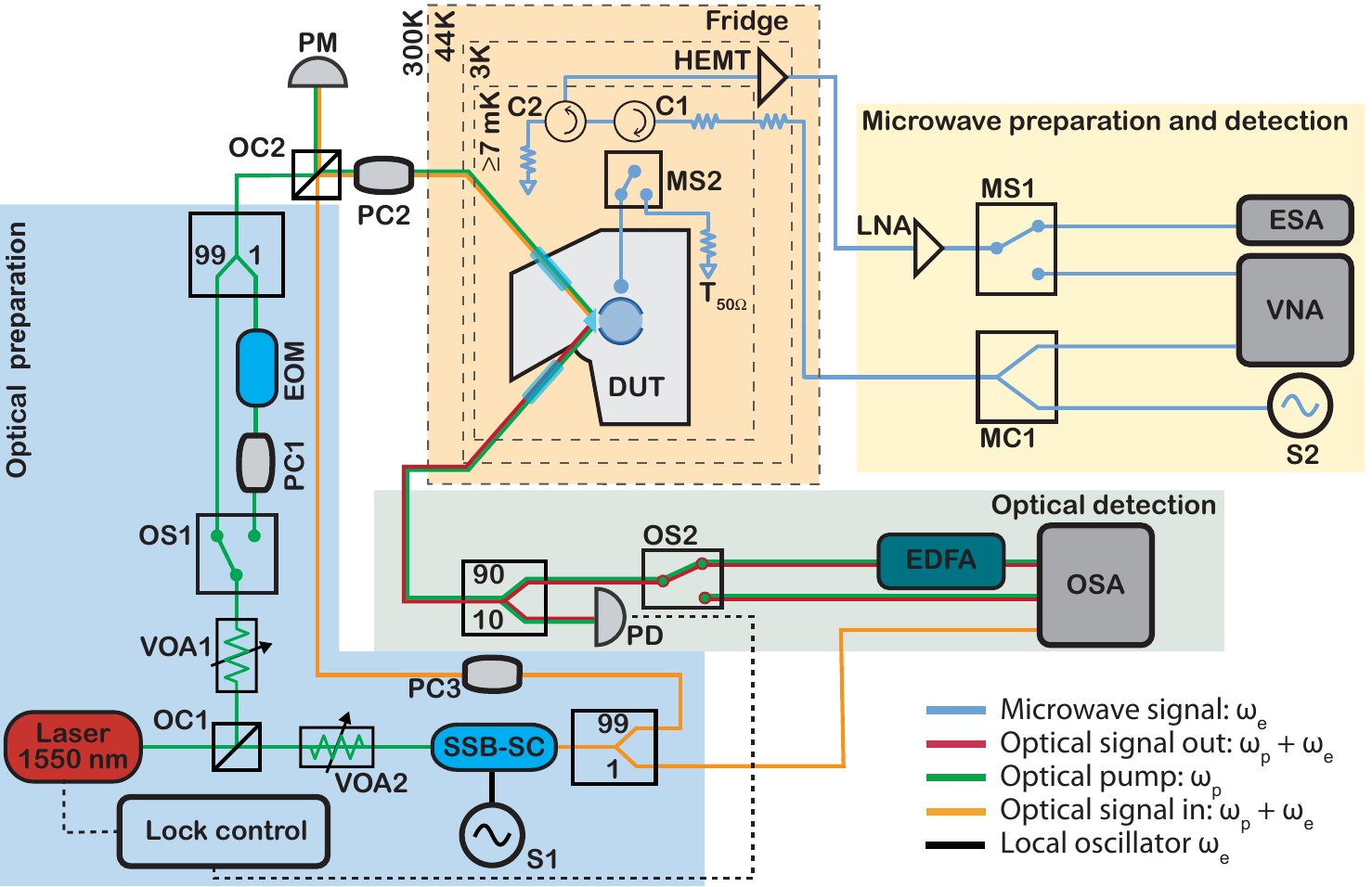}
	\caption{\textbf{Measurement setup.} A tunable laser is equally split (50/50) into two paths at the optical coupler OC1. The upper path is used as the optical pump and it goes through a variable optical attenuator VOA1 that allows to vary $P_p$. The optical pump can then be either sent directly to the cryostat fiber, or it can go first through an electro-optic modulator (EOM) in order to create sidebands for spectroscopy calibration. The second path (horizontal) is used to generate the optical signal. It goes also through a variable optical attenuator and it is then frequency up-shifted by $\omega_e$ ($\sim$FSR) using a single sideband EO-modulator with suppressed carrier (SSB-SC) driven by a microwave source with local oscillator frequency $\omega_e$ (S1). A small fraction (1\%) of this signal is picked up and sent directly to an optical spectrum analyzer (OSA) for sideband and carrier suppression ratio monitoring. The rest (99\%) is recombined with the pump at OC2, sent to the fridge input fiber and the total power is monitored with a power meter (PM). The optical tones are focused on the prism with a GRIN-lens which then feeds the WGM resonator via evanescent coupling. Polarization controllers PC2 and PC3 are set to achieve maximum coupling to a TE polarized cavity mode. The reflected (or created) optical sideband signal and the reflected pump are collected with the second GRIN-lens and coupled to the cryostat output fiber. The optical signal is then split: 90\% of the power goes to the OSA and 10\% is sent to a photodiode (PD), which is used for mode spectroscopy and to lock on the optical mode resonance during the conversion measurement. The 90\% arm is either sent directly to the OSA, or goes through an EDFA for amplification, depending on the microwave to optics converted signal power. On the microwave side, the signal is sent from the microwave source S2 (or from the VNA for microwave mode spectroscopy) to the fridge input line via the microwave combiner (MC1). The input line is attenuated with attenuators distributed between 3 K and 10 mK with a total of 60\,dB
	in order to suppress room temperature microwave noise. Circulator C1 redirects the reflected tone from the cavity to the amplified output line, while C2 redirects noise coming in from the output line to a matched 50$\Omega$ termination. The output line is amplified at 3 K by a HEMT-amplifier and then at room temperature again with a low noise amplifier (LNA). The output line is connected to switch MS1, to select between an ESA or a VNA measurement. Lastly, microwave switch MS2 allows to swap the device under test (DUT) for a temperature $T_{\text{50}\Omega}$ controllable load, which serves as a broad band noise source in order to calibrate the output line's total gain and added noise (see \ref{calibration})}
	\label{fullsetup}
\end{figure}

The measurement setup used to characterize the performance of the electro-optic converter is shown in Fig.~\ref{fullsetup}.

%

\section{Optical resonator}
\subsection{WGM resonator fabrication}
The whispering gallery mode (WGM)  resonator was manufactured from a z-cut congruent undoped lithium niobate wafer. The resonator initial dimensions were a major radius of $R=2.5$ mm, a curvature radius of $\rho\approx0.7$ mm and an initial  thickness $d = \SI{0.5}{mm}$.  The lateral surface was polished with diamond slurry from 9 $\mu$m (rms particle diameter) down to 1 $\mu$m. Subsequently, the resonator was thinned down to  a 0.15 mm thickness with 5 $\mu$m diamond slurry in a lapping machine. Top and bottom surfaces were then finished by chemical-mechanical polishing.

\subsection{Optical prism coupling}
We couple the optical pump into the resonator via frustrated total internal reflection between the prism and the resonator surface. The optical beam coming from the cryostat input optical fiber is focused to the coupling window with an angle $\Phi_c\approx50^\circ$ using a gradient index (GRIN) lens (see Fig. \ref{figure_1}a). The reflected pump and the converted optical signal are caught by a second GRIN lens and directed to the the cryostat output optical fiber. The diamond prism is an  isosceles triangle with basis angle 53$^\circ$ and height 0.8 mm. The prism's input and output sides are anti-reflexion coated, and it is fixed from the backside to a copper wire as shown in Fig. \ref{figure_1}a. The copper wire goes through a small canal outside the microwave cavity and is attached to a linear piezo-positioning stage (PS). This way the distance $d$ between the WGM resonator and the prism coupling surface can be controlled with nanometer scale precision.


In order to reduce GRIN lens misalignments during cool down, we machine a single piece, oxygen-free copper holder which has the prism-WGM resonator coupling point at its center. Furthermore, we set up a low temperature realignment system which consists of two ANPx101-LT and one ANPz101-LT PSs from attocube for each GRIN lens, allowing us to align them in the x-y-z direction. A feedback algorithm tracks the overall optical transmission as well as the optical mode contrast during the dilution refrigerator cooldown to 3 K where the final alignments are performed before condensation and further cooldown to the cryogenic base temperature of about 7 mK.

\subsection{Optical characterization}
\label{optical_spectroscopy}
\begin{figure}[t]
	\centering 
		\includegraphics[width=0.75\textwidth]{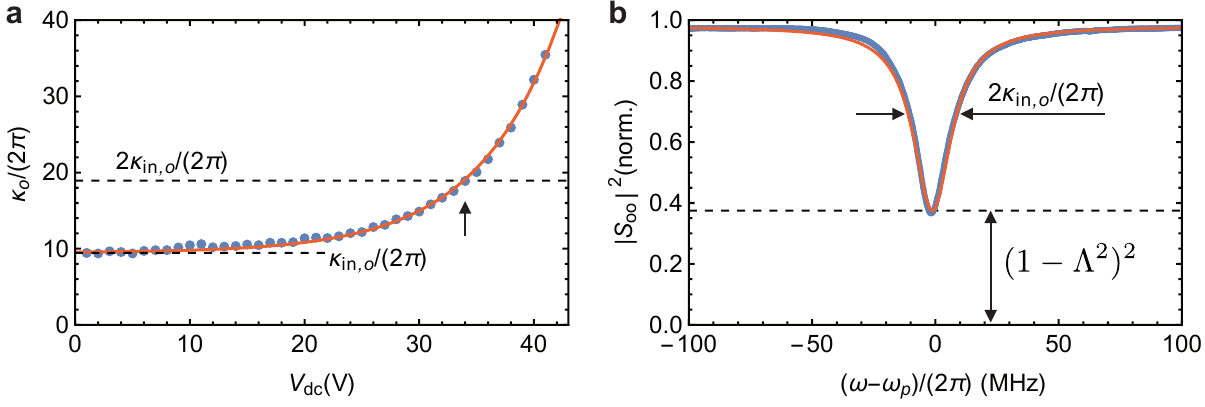}
	\caption{\textbf{Optical coupling. a,} 
	Measured $\kappa_o/(2\pi)$ as a function of piezo Voltage $V_\text{dc}$ (blue points) and exponential fit (red line). For large distances $d\propto -V_\text{dc}$ we find $\kappa_o/(2\pi)=\kappa_{\text{in},o}/(2\pi)=9.46$ MHz. The lower dashed line indicates the lower limit $\kappa_o=\kappa_{\text{in},o}$. The upper dashed line shows the critical coupling condition where $\kappa_{\text{in},o}=\kappa_{\text{ex},o}$ and $ \kappa_{o}=2\kappa_{\text{in},o}$ with the corresponding $V_\text{dc}$ (vertical arrow).
	\textbf{b,} Measured optical reflection spectrum around 1550 nm at base temperature for critical coupling (arrow in panel a). Due to imperfect optical mode matching quantified by $\Lambda^2=0.38$ an amount proportional to $(1-\Lambda^2)^2$ of the input power is reflected at critical coupling (dashed line).	}
	\label{piezotuning}
\end{figure}

The optical resonator is characterized by analyzing its reflection spectrum. We sweep the frequency $\omega/(2\pi)$ of an optical tone over several GHz around 1550 nm and measure the intensity of the reflected signal on a photodiode (PD in Fig.~\ref{fullsetup}).  In Fig.~\ref{piezotuning}a we show the pump mode spectrum for a TE polarized  tone, whose polarization is parallel to the WGM resonator's symmetry and optical axis. The optical free spectral range (FSR) for this mode was measured by superimposing it with EOM-generated sidebands from modes one FSR away \cite{Rueda2016} (see Fig.~\ref{fullsetup} for the EOM). The measured optical FSR changed from 8.79 GHz at room temperature to 8.82 GHz at base temperature.

To characterize the coupling to the optical system, we measure the optical mode spectrum for different positions of the prism, thus changing $\kappa_{\text{ex}, o}$. The normalized spectrum of the chosen mode follows the analytical model~\cite{RuedaSanchez2018}
\begin{equation}
\frac{|\text{S}_{oo}(\omega-\omega_o)|^2}{|\text{S}_{oo}(\Delta\omega)|^2}=1-\frac{4\kappa_{\text{ex},o}\Lambda^2(\kappa_o-\Lambda^2\kappa_{\text{ex},o})}{\kappa^2_o+4(\omega-\omega_o)^2}\label{outwave}, 
\end{equation}
where the factor $\Lambda$ describes the electric field overlap between the evanescent tail of the beam reflecting on the prism and the resonator mode and $\Delta\omega\gg\kappa_o$. The external coupling rate $\kappa_{\text{ex},o}$ strongly depends on the distance $d$ between the prism and the resonator
\begin{equation}
\kappa_{\text{ex},o}(d)=\kappa_{\text{ex},o}^\text{max} \cdot \exp(-k_0\cdot d)
\end{equation}
with $\kappa_{\text{ex},o}^\text{max} = \kappa_{\text{ex},o}(d=0)$ 
and $k_0=\omega_o\sqrt{n^2_\text{LN}-1}/c$ \cite{RuedaSanchez2018}, $n_\text{LN}$ the refractive index of LN and $c$ the speed of light in vacuum. We control the distance by applying a DC-voltage to the piezo stage $d\propto-V_\text{dc}$ and measure the transmission spectrum. The fitted total optical linewidth $\kappa_o=\kappa_{\text{ex},o}+\kappa_{\text{in},o}$ 
as a function of $V_\text{dc}$ is shown in  Fig.~\ref{piezotuning}a. From an exponential fit of the measured $\kappa_o$ (red line), we extract the offset corresponding to $\kappa_{\text{in},o}/(2\pi) = 9.46$ MHz. Furthermore, at critical coupling ($\kappa_{\text{ex},o}=\kappa_{\text{in},o}$), we extract $\Lambda^2=0.38$ from a fit to Eq.~(\ref{outwave}) as shown in Fig.~\ref{piezotuning}a.
The intra-cavity photon number for the optical pump is given by
 \begin{equation}n_p(\omega)=\frac{P_p\Lambda^2}{\hbar\omega_p}\cdot\frac{4\kappa_{\text{ex},o}}{\kappa_o^2+4(\omega-\omega_p)^2}\label{phnumber},\end{equation}
where $P_p$ stands for the optical pump power sent  to the resonator-prism interface. 
The WGM resonators's FSR and linewidth do not change over the full optical pump power range.

\section{Microwave cavity}
\subsection{Design}
The conversion efficiency between the optical and microwave modes depends strongly on the microwave electric field confinement at the rim of the WGM resonator. Our hybrid system, based on a 3D-microwave cavity and a WGM resonator, offers a high degree of freedom to control the microwave spatial distribution $\psi_e(\vec{r})$, microwave resonance frequency $\omega_e$ and external coupling rate $\kappa_{\text{ex},e}$. We used finite element method (FEM) simulations in order to find suitable design parameters for the microwave cavity.

\begin{figure}[t]
	\centering
		\includegraphics[width=0.8\textwidth]{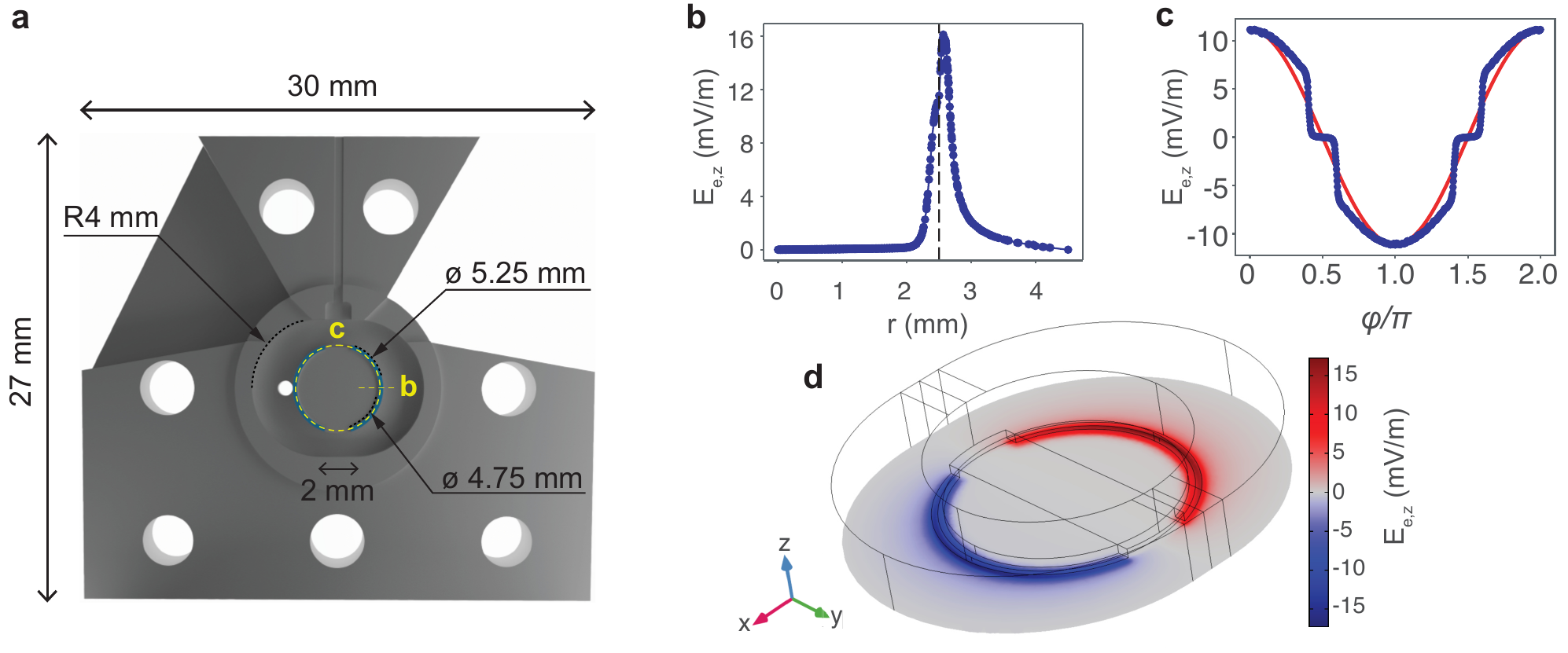}
	\caption{\textbf{Microwave cavity design.} \textbf{a,} Computer aided design drawing of the top part of the aluminum microwave cavity. \textbf{b,} FEM simulation of the single photon electric field distribution of the $m=1$ mode of the microwave cavity. Shown is the z direction of the field along the radial axis at $z=0$ (in the center of the WGM) and $\phi=0$ (yellow dashed line in panel a). The horizontal dashed line marks the position of the edge of the WGM resonator. \textbf{c,} FEM simulation of the z component of the single photon electric field taken at the position of the optical mode maximum and plotted as a function of the azimuthal angle $\phi$ (yellow dashed circle in panel a). The red line is a sinusoidal function as a guide to the eye. \textbf{d,} FEM simulation of the z component of the single photon electric field in the $z=0$ plane of the $m=1$ microwave resonance.}
	\label{setupsim}
\end{figure}

A schematic drawing of the microwave cavity with its important dimensions is shown in Fig.~\ref{setupsim}a. The LiNbO$_3$ WGM resonator is clamped between two aluminum rings (highlighted in blue). In this way we maximize the microwave electric field overlap with the optical mode, the latter being confined close to the rim of the WGM resonator (see Fig.~\ref{setupsim}b). The microwave spatial electric field distribution shows one full oscillation along the circumference of the WGM resonator (see Fig.~\ref{setupsim}c and d) to fulfill the phase matching condition. The aluminum rings have a cut in the middle in order to maximize the field participation factor and minimize potential magnetic losses in the dielectric. The cavity's cylindrical inner volume can be tailored to achieve the desired microwave resonance frequency, which can then be tuned by $\sim500$\,MHz \textit{in situ}, by moving an aluminum cylinder placed inside the lower ring. This allows to compensate the thermal contraction induced frequency shift that occurs during cooldown of the device. The top right part of the shown top half of the cavity is cut out in order to facilitate the assembly of the device.

\subsection{FEM simulation of electro-optic coupling}
From FEM simulations we obtain the single photon spatial electric field distribution given as $E_{e,z}(r, z, \phi) = E_{e,z}^ \text{max}\Psi_e(r,z)(1+f(\phi))\cos(\phi)$, where $\Psi_e(r,z)$ is normalized to 1, $r = \sqrt{x^2+y^2}$ and $\phi = \text{arctan(y/x)}$, see also Fig \ref{setupsim}d. For this simulation we used the reported \cite{wong2002properties} dielectric permittivity of lithium niobate at 9~GHz, i.e.~$\varepsilon_{e}={(42.5, 42.5, 26)}$.
The function $f(\phi)$ is symmetric and describes the deviation of the azimuthal field distribution from a pure sinusoidal shape as shown in Fig.~\ref{setupsim}c. The optical mode is distributed along the ring $\{r=r_o,\; z=0,\; \phi\in [0\;2\pi]\}$ and the maximum value of the microwave electric field on this ring is $E_{eo}=E_{e,z}(r_o,0,\phi_\text{max})=E_{e,z}^ \text{max}\Psi_e(r_o,0)=11.1$\,mV/m. The optical mode being a clockwise (C) traveling wave, we must decompose the stationary microwave field into a clockwise and a counterclockwise (CC) traveling wave in order to calculate the coupling
\begin{eqnarray}
E_{e,z}(r_o, 0, \phi, t) &=& E_{eo}(1+f(\phi))\cos(\phi)\cos(\omega_e t)\nonumber\\
&=&E_{eo}(1+f(\phi))\left(\frac{e^{-i(\phi- \omega_e t)}+e^{i(\phi- \omega_e t)}+e^{-i(\phi+\omega_e t)}+e^{i(\phi+ \omega_e t)}}{4}    \right)\nonumber\\  
&=&E^+_\text{C}+E_\text{C}^-+E^+_\text{CC}+E^-_\text{CC}   . 
\end{eqnarray}
By introducing $E_{e,z}(r_o,0,\phi,t)$ into Eq.~(\ref{g_formula}), we get ($E^+_\text{CC}$ and $E^-_\text{CC} $ do not participate in the interaction)
\begin{equation}
g=n_pn_or  \sqrt{\frac{\omega_p\omega_o}{4V_oV_p}} \frac{E_{eo}}{4}\int dV( \psi_o^* \psi_pe^{-i\phi}+\psi_o^* \psi_pe^{-i\phi}f(\phi)).\label{defgg}
\end{equation}
Where $n_p$ and $n_o$ are the refractive indices of the pump $\omega_p$ and the sideband $\omega_o$. The effective mode volumes $V_k$ are given by the integral $\int\text{d}V\psi_k\psi_k^*$ over the respective optical field spatial distributions $\psi_p=\Psi_p(r,\theta)e^{-im}$ and $\psi_o=\Psi_o(r,\theta)e^{-i (m+1)}$. The second term in the integral in Eq.~(\ref{defgg}) is zero due to the symmetry of $f(\phi)$, reducing Eq.~(\ref{defgg}) to
\begin{equation}
g=\frac{1}{8}\ n_p^2  \omega_o  r_{33}  E_{eo} \label{qwithtroot}
\end{equation}
where $n_o\approx n_p=2.13$ ($\varepsilon_o\approx \varepsilon_p=4.54$) is the extra-ordinary refractive index (dielectric permittivity) of LN at $\omega_o\approx\omega_p=(2\pi)\times193.5$ THz and $r_{33}=31$ pm/V is the electro-optic coefficient. For these values we estimate $g_\text{sim}/(2\pi)=38$ Hz at room temperature. 

\subsection{Room temperature measurement of $g$}
\begin{figure}[h]
\centering
\includegraphics[width=0.4\textwidth]{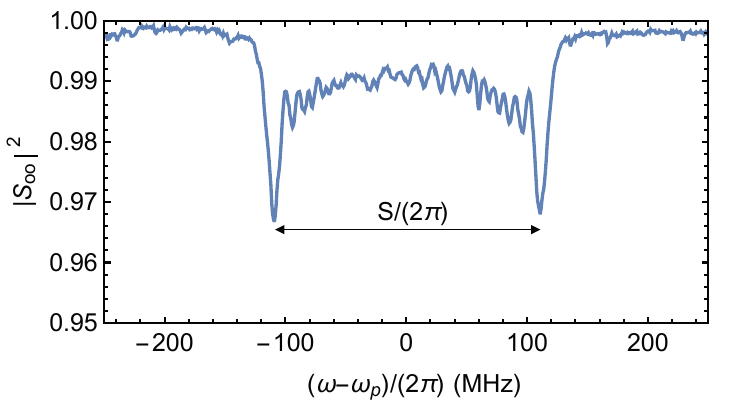}
\caption{\textbf{Direct measurement of $g$.} Measured optical mode splitting of $S/(2\pi)= 220$\,MHz at room temperature, obtained with an input power of 9.3\,dBm applied to the microwave cavity at $\omega_e$. $S \approx 4 \sqrt{n_e} g_\text{rt} $ is related to the measured single photon coupling strength $g_\text{rt}/(2\pi)=36.1$\,Hz via the calculated intra-cavity photon number $n_e=2.3\times10^{12}$  \cite{Ruedacombs}.}
\label{tuning_micro-cry}
\end{figure}

The system was assembled at room temperature and a microwave tone was fed into the cavity with a coaxial probe coupler of length 1.2 mm. By displacing the tuning cylinder the cavity frequency $\omega_e/(2\pi)$ could be shifted from 8.40 to 9.22 GHz, slightly shifted up compared to numerical simulations. We attribute this to small air gaps between the WGM resonator and the aluminum disk, which decrease the effective dielectric constant between the electrodes. To match the measured frequency range exactly, we introduce an air gap of only $\sim$ 1 $\mu$m in the simulations, bringing down the estimated coupling to $g_\text{sim}/(2\pi)=36.2$\,Hz, 

At $\omega_e=\text{FSR}$, the microwave mode has the parameters  $\kappa_{\text{ex},e}/(2\pi) =  2.48$\,MHz and $\kappa_{\text{in},e}/(2\pi) = 29$\,MHz. We infer the nonlinear coupling constant of the system by applying a strong microwave drive tone to the cavity and measuring the resulting optical mode splitting $S \approx 4 \sqrt{n_e} g_\text{rt} $ as described in Ref.~\cite{Ruedacombs}. In Fig.~\ref{tuning_micro-cry}a we show a measured splitting of $S/(2\pi)=220$\,MHz for a 9.3\,dBm microwave pump power applied on resonance. This corresponds to $g_\text{rt}/(2\pi)=36.1$\,Hz, a five fold improvement compared to earlier results \cite{Rueda2016}, and in excellent agreement with the simulations. 

\subsection{Microwave cavity fabrication}
The microwave cavity is milled out of a block of pure aluminum (5N). It is divided into a lower and upper part, that are closed after placing the WGM resonator and the prism using brass screws. The internal geometry can be seen in Fig. \ref{setupsim}a. When closing the cavity the pure aluminum rings get in contact with the optical resonator. The rings deform slightly, which minimizes the formation of air gaps that would otherwise reduce $g$.
 
\subsection{Microwave characterization}
\label{mw_spectroscopy}
The microwave resonance tuning range was measured with a VNA connected to the cryostat transmission line as shown in Fig. \ref{fullsetup}. The resonance frequency can be tuned from  8.70 GHz to 9.19 GHz as shown in the main text. This range is at slightly higher frequency compared to the room temperature one. This we attribute to thermal contraction that leads to small air gaps. The decay rates of the microwave mode at the cryogenic base temperature and the lowest optical input power are $\kappa_{\text{ex},e}/(2\pi)= 3.7$ MHz and $\kappa_{\text{in},e}/(2\pi)=6.7$ MHz.

\begin{figure}[t]
	\centering
		\includegraphics[width=0.8\textwidth]{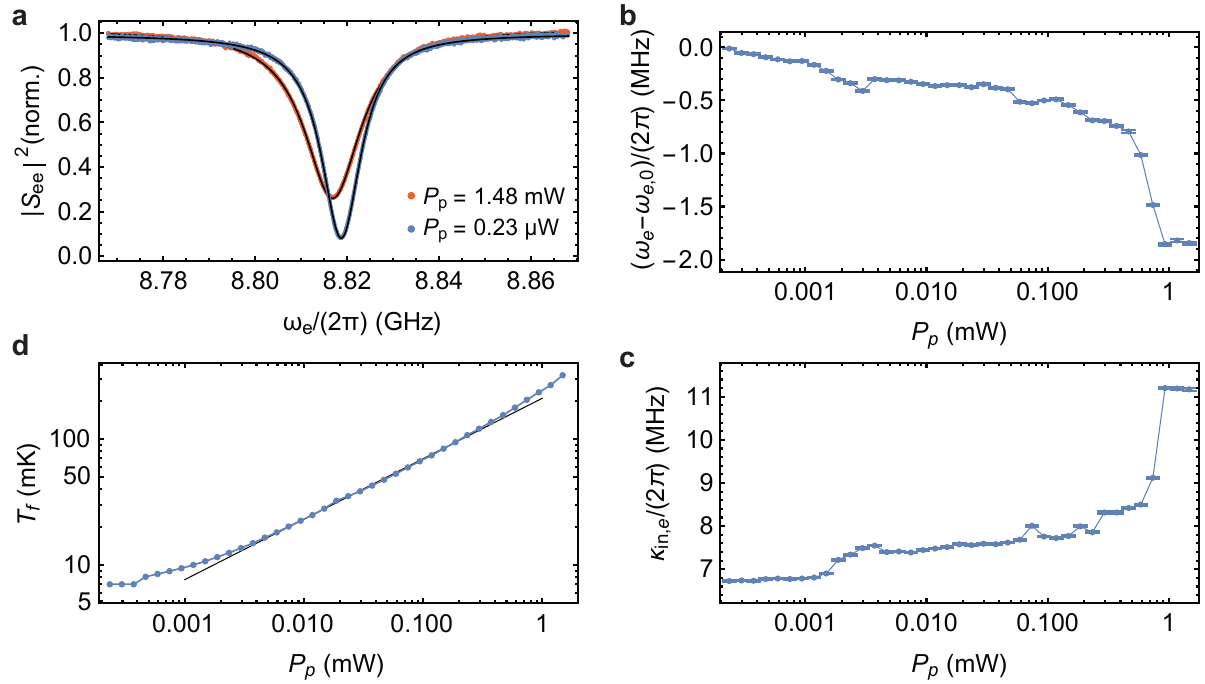}
	\caption{\textbf{Microwave cavity properties. a,} Measured reflection spectra of the microwave resonance for the minimum (blue), and maximum (red) applied optical pump power $P_p$ together with Lorentzian fits (black lines). 
	\textbf{b,} Fitted microwave resonance frequency $\omega_e-\omega_{e,0}$ as a function of $P_p$ with the error bars indicating the 95\% confidence interval of the fit. $\omega_{e,0}$ is the fitted resonance frequency obtained for the minimum optical pump power of $P_p=0.23\,\mu$W.
	\textbf{c,} Fitted microwave intrinsic loss rate $\kappa_{\text{in},e}$ as a function of $P_p$ with the error bars indicating the 95\% confidence interval of the fit.
\textbf{d,} The mixing chamber temperature sensor reading $T_f$ of the dilution refrigerator as a function of the optical pump power $P_p$ (blue points) and a power law fit $T_f\propto P_p^{0.48}$ to the intermediate power range (black line).
} \label{microwavevspump}
\end{figure}

Unlike the optical system, the microwave cavity's parameters undergo changes as a function of the optical pump power $P_p$ applied to the WGM resonator. In Fig.~\ref{microwavevspump}a we show the normalized spectra of the microwave resonance at the smallest (blue) and largest (red) optical pump power together with a Lorentzian fit. From these measurements we extract the microwave resonance frequency $\omega_e$ (shown in panel b) and the internal and external loss rates $\kappa_{\text{in,ex},e}$ (shown in panel c) as function of $P_p$. $\kappa_{\text{ex},e}$ depends only on the fixed geometry and is approximately constant. In contrast $\kappa_{\text{in},e}$ increases and $\omega_e$ decreases with increasing $P_p$ until the microwave cavity undergoes the superconducting phase transition. Once the normal conducting state is reached, a further increase of $P_p$ does not lead to any perceptible change, as can be seen in Fig.~\ref{microwavevspump} panels b and c. 
The microwave resonance red shift (see Fig. \ref{microwavevspump}c) and the $\kappa_{\text{in},e}$ increase are expected due to optically induced creation of quasiparticles in the aluminum cavity as discussed for example in Ref.~\cite{Witmer2020}. While the local heating is significant, the temperature of the mixing chamber plate of the dilution refrigerator follows a slow ($P_p^{0.48}$ at intermediate powers) rise from $T_f=7$\,mK up to $T_f=320$\,mK as shown in Fig.~\ref{microwavevspump}d.

\section{Frequency conversion}
\subsection{Theoretical model}
\begin{figure}
	\centering
		\includegraphics[width=0.8\textwidth]{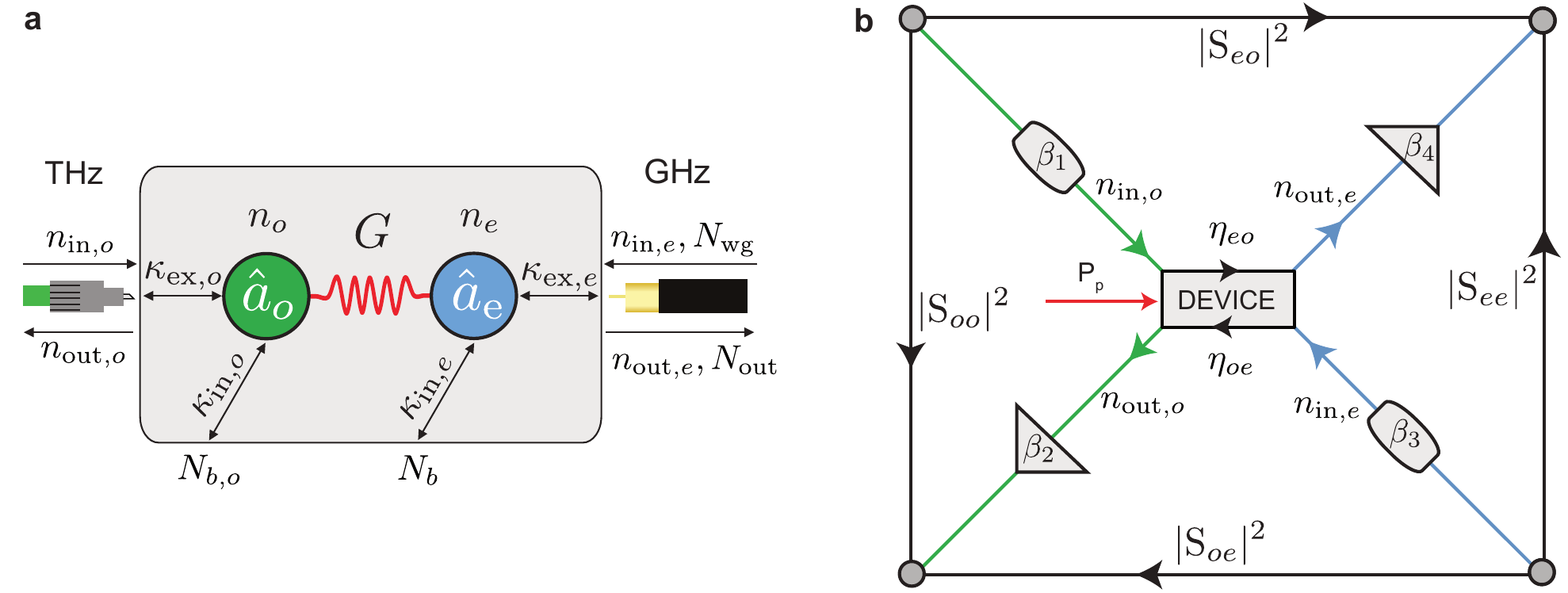}
	\caption{\textbf{Electro-optic photon conversion.} 
	\textbf{a,} Schematic representation of the microwave $(\hat{a}_e)$ and optical $(\hat{a}_o)$ modes with coherent populations $n_e$ and $n_o$ and incoherent occupancies $N_e$ and $N_o$ (not shown). The electro-optic coupling strength $G=\sqrt{n_p} g$ originates from the Pockels effect in the lithium niobate WGM resonator. Both modes are coupled to an internal thermal bath $N_b$ and $N_{b,o}$ with the rates $\kappa_{\text{in},i}$. They are also coupled the respective coaxial and fiber waveguides with the rates $\kappa_{\text{ex},i}$. At finite temperature the microwave waveguide has a thermal bath occupancy $N_\text{wg}$ and also hosts the total output noise $N_\text{out}$ measured in the experiment.
\textbf{b,} The $\text{S}_{ij}$ coefficients are defined between the microwave and optical input and output ports outside the cryostat (gray circles). Attenuation for the optical input and output path are $\beta_1=-4.81$\,dB and $\beta_2=-5.5$\,dB (without EFDA), or the gain $\beta_2=+30.8$\,dB (with EFDA). An observed gain saturation at high pump powers is measured and corrected. Attenuation and gain of the microwave input and output path are $\beta_3=-74.92$dB and $\beta_4=+67.05$dB, respectively. From two conversion measurements on resonance ($S_{ij}$) and two reflection measurements  out of resonance ($S_{ii}$) where $\eta_{ii}=1$, we infer the bidirectional photon conversion efficiency of the device according to Eq.~(\ref{conversion_autonorm}) and referenced to the input and output photon numbers $n_{\text{in},i}$  and $n_{\text{out},i}$.}
	\label{4_port_calibration}
\end{figure}

We model the input-output response of the electro-optic system by taking into account external coupling rates $\kappa_{\text{ex},i}$ and internal loss rates $\kappa_{\text{in},i}$ to the external and internal thermal baths, as shown in Fig.~\ref{4_port_calibration}a. First, we define the coherent conversion matrix 
as the ratio between the output and the input photon numbers in the absence of noise 
\begin{equation}
\eta_{ij}=n_{\text{out}, i}/n_{\text{in}, j},
\end{equation} 
for $i,j \in\{o,e\}$. The matrix $\eta_{ij}$ is derived in Ref.~\cite{Tsang2011,RuedaSanchez2018} and its explicit form is given as
\small
\begin{equation}
 \left[\begin{array}{c} n_{\text{out},o}(\omega)  \\ n_{\text{out},e}(\omega)  \end{array} \right]
 =  M(\omega)
 \begin{bmatrix}
|(i\omega+\Lambda^2\kappa_{\text{ex},o}-\kappa_o/2)(\kappa_e/2-i\omega)+|G|^2|^2 &\Lambda^2\kappa_{\text{ex},e}\kappa_{\text{ex},o}|G|^2\\
\kappa_{\text{ex},e}\kappa_{\text{ex},o}|G|^2\Lambda^2
&  |(i\omega+\kappa_{\text{ex},e}-\kappa_e/2)(\kappa_o/2-i\omega)+|G|^2 |^2
\end{bmatrix}
 \times  
 \left[ \begin{array}{c} n_{\text{in},o}(\omega)\\  n_{\text{in},e}(\omega) \end{array} \right],
\end{equation}
\normalsize
where $M^{-1}(\omega)= |(-i\omega+\kappa_o/2)(-i\omega+\kappa_e/2)+|G|^2|^2$ and $G=\sqrt{n_p}g$, with $n_p$ given by Eq.~(\ref{phnumber}).

The noise performance is one of the most important characteristics of a quantum converter. In our system, we consider two noise sources that affect the microwave mode. The first one is the noise in the waveguide $N_\text{wg}$, which can be seen as the external bath of the 50\,$\Omega$ semirigid copper coaxial port. The second noise source is given by the internal bath of the system $N_b$ as shown in Fig.~\ref{4_port_calibration}a. The optical waveguide noise $N_{\text{wg},o}$ in the fiber and the internal optical bath noise $N_{b,o}$ are neglected, because $\hbar\omega_o \ggg k_BT$. In the low cooperativity limit we also neglect the thermal occupancy of the optical mode $N_o$, which would give rise to optical output noise $N_{\text{out},o}$. 

We define the noise conversion matrix $\sigma_{ij}$ as the ratio between the output noises and the microwave input noise to the system in the absence of any coherent signal as \cite{Tsang2011,RuedaSanchez2018}
\begin{equation} \label{matnoise2}
 \left[ \begin{array}{c} N_{\text{out},o}(\omega)  \\ N_\text{out}(\omega)  \end{array} \right]=M(\omega)\begin{bmatrix}
\kappa_{\text{ex},e}\kappa_{\text{ex},o}|G|^2&\kappa_{\text{in},e}\kappa_{\text{ex},o}|G|^2  \\
|(i\omega+\kappa_{\text{ex},e}-\kappa_e/2)(-i\omega+\kappa_o/2)+|G|^2|^2 &\kappa_{\text{in},e}\kappa_{\text{ex},e}|(-i\omega+\kappa_o/2)|^2
\end{bmatrix} \times  \left[ \begin{array}{c} N_\text{wg}(\omega)\\ N_b(\omega) \end{array} \right],
\end{equation}
where $N_\text{wg}(\omega)=(\exp(\hbar\omega/k_BT_\text{wg})-1)^{-1}$ and $N_b(\omega)=(\exp(\hbar\omega/k_BT_\text{b})-1)^{-1}$ are wide band distributions compared to $\kappa_{\text{ex},e}$, such that they can be approximated as constant. 

The full input-output model including vacuum noise is given as
\begin{equation}
\textbf{n}_\text{out}(\omega)+\textbf{N}_\text{out}(\omega)=\eta_{ij}(\omega)\cdot \textbf{n}_\text{in}(\omega)+\sigma_{ij}(\omega)\cdot \textbf{N}_{\{\text{wg}, b\}}(\omega)+\textbf{n}_\text{vac}. \label{FDC}
\end{equation}
with $\textbf{n}_\text{vac}=\left[ \begin{array}{c} 0.5\\ 0.5 \end{array} \right]$.

The device is fixed to the mixing chamber of a dilution refrigerator with a base temperature of $\sim$ 7mK, preventing direct access to the device's input and output ports, see Fig.~\ref{fullsetup}. In Fig.~\ref{4_port_calibration}b we present a simplified schematic of the measurement setup, with attenuation and gain $\beta_1$, $\beta_2$ for the optical path, and $\beta_3$, $\beta_4$ for the microwave path. We define the measured scattering matrix including the transmission lines on resonance as
\begin{equation}
|\text{S}_{ij}(\omega_0)|^2=\frac{1}{(1+C)^2}
\begin{bmatrix}
\beta_2(2\Lambda^2\eta_o-1+C)^2 \beta_1 &\Lambda^2\beta_24\eta_o\eta_e C \beta_3\\
\beta_44\eta_o\eta_e C \beta_1\Lambda^2& \beta_4 (2\eta_{e}-1+C)^2 \beta_3 
\end{bmatrix},
\end{equation}
where $\eta_i=\kappa_{\text{ex},i}/\kappa_i$ and $C=\frac{4 n_p g^2}{\kappa_o\kappa_e}$ stands for the multi-photon electro-optic cooperativiy. For large signal detuning $\omega_\Delta=\omega-\omega_0\gg\kappa_e,\kappa_o$ with $\omega_0=\omega_e, \omega_o$ the scattering matrix simplifies to
\begin{equation}
|\text{S}_{ij}(\omega_\Delta)|^2= 
\begin{bmatrix}
\beta_2\beta_1 &0\\
0& \beta_4 \beta_3 
\end{bmatrix}.
\end{equation}

We infer the bidirectional conversion efficiency at each optical pump power by measuring the microwave-to-optics and optics-to-microwave transmissions on resonance and the microwave-to-microwave and optics-to-optics reflections off resonance. The total device efficiency can then be defined as
\begin{equation}
\eta_\text{tot}=\sqrt{ \frac{|\text{S}_{eo}(\omega_0)|^2 \cdot |\text{S}_{oe}(\omega_0)|^2}{|\text{S}_{ee}(\omega_\Delta)  |^2 \cdot |   \text{S}_{oo}(\omega_\Delta)   |^2 } }= \eta_o\eta_e \Lambda^2\frac{ 4C}{(1+C)^2}. \label{deftrans}
\end{equation}
In the limits for $C\ll1$ this can be approximated as
\begin{equation}
\eta_\text{ext} \approx 4\eta_o\eta_e \Lambda^2 C=\frac{64\eta^2_o\eta_e\Lambda^4g^2P_p}{\hbar\omega_p\kappa_o^2\kappa_e}. \label{deftrans}
\end{equation}
This equation was used to calculate the nonlinear coupling constant in the main text. It can be also shown that in the limit of $C\ll1$, $\Lambda<1$ and $\eta_e\ne0.5$, the bidirectional efficiency can be estimated using only resonant measurements
\begin{equation}
\eta_\text{tot}=(2\Lambda^2\eta_o-1)(2\eta_e-1)\sqrt{ \frac{     |\text{S}_{e o}(\omega_0)|^2 \cdot |\text{S}_{oe}(\omega_0)|^2    }{     |\text{S}_{ee}(\omega_0)   |^2 \cdot |  \text{S}_{oo}(\omega_0)|^2}     } = \eta_o\eta_e \Lambda^2\frac{4 C}{(1+C)^2},
\end{equation} 
where $\Lambda$ and $\eta_i$ are measured accurately from microwave and optical spectroscopy.

The system noise originates from the microwave resonator and waveguide baths $N_b$ and $N_\text{wg}$ respectively. By applying the matrix to the noise vector in Eq.~(\ref{matnoise2}) we can solve for the output noise $N_\text{out}$, which simplifies in the low cooperativity limit ($G^2\ll \kappa_o \kappa_e$) to
\begin{equation}
N_{\text{out},e}(\omega)=  \frac{4\kappa_{\text{in},e}\kappa_{\text{ex},e}}{\kappa^2_e+4\omega^2}\left(N_b-N_\text{wg}\right)   + N_\text{wg}+ 0.5.
\end{equation}
In our system the resonator bath $N_b$ is always hotter than the waveguide bath $N_\text{wg}$, because the dominant part of the dielectric absorption takes place right inside the resonator. Therefore, the output noise spectrum $N_\text{out}(\omega)$ always consists of a Lorentzian function with amplitude $N_b-N_\text{wg}$ on top of the broad band noise level $N_\text{wg}$ as shown in Fig.~\ref{fig_3}. Finally, following the same formalism the integrated (dimensionless) internal microwave mode occupancy is given as
\begin{equation}
N_e =  \frac{\kappa_{\text{ex},e}N_\text{wg}+\kappa_{\text{in},e}N_b}{\kappa_e} = \eta_eN_\text{wg} + (1-\eta_e)N_b.
\end{equation}

\subsection{Microwave calibration}\label{calibration}
\begin{figure}[b]
\centering
		\includegraphics[width=0.37\textwidth]{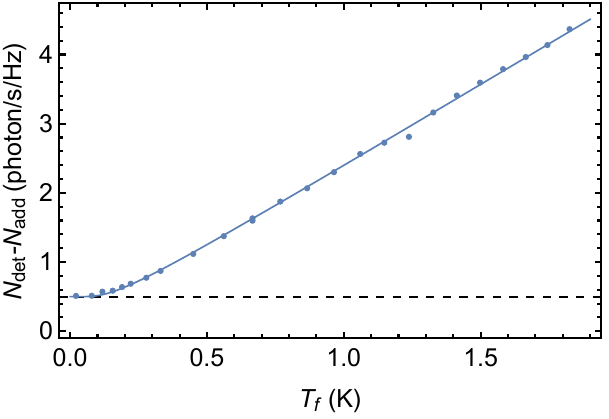}
	\caption{\textbf{System noise calibration.} Measured noise (dots) together with a fit to Eq.~(\ref{50ohmPSD}) (line) shown in units of photons using $N_\text{det}-N_\text{add}=P_\text{ESA}/(\hbar\omega_e\beta_4 \text{BW})-N_\text{add}$. The dashed line indicates the half vacuum noise photon added by the measurement.}
	\label{calibrationC}
\end{figure} 
The microwave  transmission line is characterized by the input attenuation $\beta_3$, the output gain $\beta_4$ and the total added noise of the output line $N_\text{sys}$. The output line is first calibrated by using a 50~$\Omega$ load, a resistive heater, and a thermometer that are thermally connected. Weak thermal contact to the mixing chamber of the dilution refrigerator allows to change the temperature $T_{50\Omega}$ of the 50~$\Omega$ load without heating up the mixing chamber. We vary $T_{50\Omega}$ from 21.5 mK to 1.8 K and measure the amplified thermal noise on a spectrum analyzer. 
The measured power spectral density $P_\text{ESA}(\omega)$ is approximately constant around the microwave resonance frequency $\omega_e$ and its temperature dependence follows
\begin{equation}
P_\text{ESA} = \hbar\omega_e\beta_4 \text{BW}\left(\frac{1}{2} \coth{\left(\frac{\hbar\omega_e}{2 k_\text{B}T_{50\Omega}}\right)}+N_\text{add}\right)
\label{50ohmPSD},
\end{equation}
where $\text{BW}$ stands for the chosen resolution bandwidth, $k_\text{B}$ is Boltzmann's constant and $N_\text{add}$ is the effective noise added to the signal at the output port of the device due to amplifiers and losses. At $T_{50\Omega}=0$\,K this reduces to $P_\text{ESA} = \hbar\omega_e\beta_4 \text{BW}N_\text{sys}$ with $N_\text{sys}=N_\text{add}+0.5$. 
Figure~\ref{calibrationC} shows the detected noise $N_\text{det}-N_\text{add}=P_\text{ESA}/(\hbar\omega_e\beta_4 \text{BW})-N_\text{add}$ as a function of the load temperature $T_{50\Omega}$.
The values for gain and added noise obtained from a fit to Eq.~(\ref{50ohmPSD}) are 
$67.65\pm 0.05$\,dB and 
$10.66\pm 0.15$ as shown in Fig.~\ref{calibrationC}. The emitted black body radiation undergoes the same losses and gains, as shown in Fig.~\ref{fullsetup}, up an independently calibrated cable length difference right at the sample output resulting in an additional loss of $0.6\pm 0.09$\,dB. Taking into account this addition loss we arrive at the corrected gain and system noise $\beta_4=67.05\pm 0.16$\,dB and $N_\text{sys}= 12.74\pm 0.36$.
For the stated error bars we take into account the 95\% confidence interval of the fit, an estimated temperature sensor accuracy of $\pm 2.5\%$ over the relevant range, as well as the estimated inaccuracy in the cable attenuation difference. The input attenuation is then easily deduced from a VNA reflection measurement $|\text{S}_{ee}|^2$ that yields $\beta_3=-74.92\pm0.16$\,dB.


\subsection{Optical calibration}
The optical transmission lines consist mainly of two optical single mode fibers. The input optical line starts from the OC2 (see Fig.~\ref{fullsetup}) and terminates at the WGM resonator-prism interface. The output optical line is defined from the WGM resonator-prism interface to the OSA (see Fig.~\ref{fullsetup}). From the measured external conversion efficiencies $\eta_\text{tot}$ and the microwave line calibration, we can determine the losses of the input and output transmission lines using
\begin{eqnarray}
\frac{P_{\text{out}, o}}{\omega_o}=\beta_2\eta_\text{tot}\beta_3 \frac{P_{\text{in}, e}}{\omega_e}\nonumber\\
\frac{P_{\text{out}, e}}{\omega_e}=\beta_4\eta_\text{tot}\beta_1 \frac{P_{\text{in}, o}}{\omega_o}
\label{powerratios}
\end{eqnarray}
where $P_{\text{in}, i}$ are the input powers  of our transmission lines coming out from OC2 and S2 and $P_{\text{out}, i}$ are the measured powers at the end of the transmission lines measured with the OSA and ESA. The procedure yields the input attenuation $\beta_1=-4.81$ dB and the output gain (via EDFA) $\beta_2=+30.8$ dB. For measurements above $P_p=0.1$ mW, we bypass the EDFA by switching OS2, resulting in an output attenuation of $\beta_2=-5.5$dB. 

\subsection{Bidirectionality}
\begin{figure}[h]
\centering
		\includegraphics[width=0.40\textwidth]{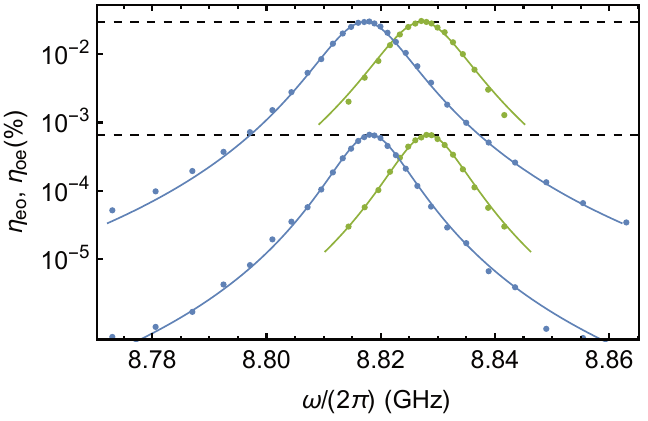}
	\caption{\textbf{Conversion bandwidth and bidirectionality.} Measured total conversion efficiency $\eta_{eo}$ (blue dots) and $\eta_{oe}$ (green dots) as a function of output signal frequency $\omega$ (optics to microwave, blue) or $\omega=\omega_o-\omega_p$ (microwave to optics, green) for $P_p$=18.7\,$\mu$W and $P_p$=1.48\,mW. Microwave-to-optics conversion (green) is displaced by 10\,MHz for better visibility. Dashed lines indicate the maxima of the theory curves in agreement with the results reported in the main text.}
	\label{bidirectional}
\end{figure}

Figure \ref{bidirectional} shows the measured total conversion efficiency as a function of signal frequency (dots) using Eq.~(\ref{powerratios}) together with theory (lines) using Eq.~(\ref{Bandwidth}) in both conversion directions for two different pump powers. Because the optical calibration Eq.~(\ref{powerratios}) assumes symmetric bidirectionality we also find that the measurement results are perfectly symmetric. Nevertheless, direct measurements of $\beta_1$ taken at room temperature of -2.6~dB are in good agreement with the optical calibration. We attribute the additional loss of up to 2.2~dB to changes in the optical alignment during the cooldown, e.g.~in the cold APC connector, as well as reflection loss at the first prism surface that is not included in the room temperature calibration. 

\end{document}